\documentclass[epj]{svjour}
\usepackage{latexsym}
\usepackage[dvips,dvipdf]{graphicx}
\usepackage{epsf}

\begin{document}

\title{Electromagnetic form factor via
Bethe-Salpeter amplitude in Minkowski space}
\author{ J. Carbonell\inst{1}, V.A. Karmanov\inst{2}, \and M. Mangin-Brinet\inst{1}
} \institute{Laboratoire de Physique Subatomique et Cosmologie, 53
avenue des Martyrs, 38026 Grenoble, France \and Lebedev Physical
Institute, Leninsky Prospekt 53, 119991 Moscow, Russia}
\date{}
\abstract{For a relativistic system of two scalar particles, we
find the Bethe-Salpeter amplitude in Minkowski space and use it to
compute the electromagnetic form factor.  The comparison with
Euclidean space calculation shows that the Wick rotation in the
form factor integral induces errors which increase with the momentum
transfer $Q^2$. At JLab domain ($Q^2=10\; GeV^2/c^2$), they are about
30\%. Static approximation results in an additional and more significant
error. On the contrary, the form factor calculated in light-front dynamics is
almost indistinguishable from the Minkowski space one.
} 

\authorrunning{J. Carbonell \and V.A. Karmanov \and M. Mangin-Brinet}
\titlerunning{Electromagnetic form factor via
Bethe-Salpeter amplitude...}
 \maketitle
\bibliographystyle{unsrt}

\section{Introduction}
The Bethe-Salpeter (BS) equation \cite{SB_51} provides a
field-theoretical framework for a relativistic treatment of
few-body systems. It has been extensively studied in the
literature (see \cite{nakanishi} for a review) and used to obtain
relativistic descriptions of two-body bound and scattering states.

BS equation is naturally formulated in the momentum
representation. In Minkowski space, for two spinless particles, it
reads:
\begin{eqnarray}\label{bs}
\Phi_M(k;p)&=&\frac{i^2}{\left[(\frac{p}{2}+k)^2-m^2+i\epsilon\right]
\left[(\frac{p}{2}-k)^2-m^2+i\epsilon\right]}
\nonumber\\
&\times&\int \frac{d^4k'}{(2\pi)^4}iK_M(k,k',p)\Phi_M(k';p)
\end{eqnarray}
The interaction kernel $K_M$ is given by irreducible Feynman
diagrams. Any finite set of them is an approximation of the
interaction Lagrangian of the theory under consideration. Most of
works were done in the ladder approximation, that is restricting
the interaction kernel to its lowest order exchange term. Several
researches (in particular \cite{NT_PRL_96}) indicate that the
higher order kernels, usually not incorporated in the BS equation,
give a significant contribution to the two-body binding energy.
Studies of these contributions, step by step, in the BS framework,
would be of indubitable interest.

Until very recently, the BS equation had been solved only in the
Euclidean space, {\it i.e.} after performing a Wick rotation
\cite{W_54}, in order to remove the singularities due to the free
propagators. The validity of Wick rotation has been proved in
\cite{W_54} for the ladder kernel. For higher order kernels ({\it
e.g.}, for the cross box) the  possibility of Wick rotation is
less clear, since one deals with the "partially Euclidean" BS
amplitude,  in which the relative energy $k_0= ik_4$ is
imaginary, whereas the total energy  $p_0$ remains real. This
"partially Euclidean" transformation is a subtle point to be
checked more carefully. We will see below that it is valid also
for the cross-ladder kernel. Then the Euclidean BS equation (in
the rest frame) provides exactly the same binding energy as the
Minkowski one.

If we are interested not only in the binding energy but also in
the electromagnetic (EM) form factors, the Euclidean BS amplitude
in the rest frame is not enough. On one hand, when computing the
integral for the form factor, the rotated contour crosses
singularities of the integrand. So, the result is not reduced to
the naive replacement $k_0=ik_4$ which transforms the Minkowski BS
amplitude into the Euclidean one. On the other hand, form factor
for non-zero momentum transfer involves the BS amplitude for
non-zero total momentum $p$. This amplitude can be obtained from
the rest frame one by a boost, but the parameters of this boost
for real $p$ and imaginary $k_0$ are complex. This requires the
knowledge of the BS amplitude in the full complex plane. The
continuation of the Euclidean amplitude from real axis to the
complex plane is numerically very unstable and can hardly be done
in practice. Instead of it, the BS amplitude in the complex plane
can be found by solving the equation for complex arguments.
However, the equation in the complex plane is no longer Euclidean
nor Minkowski one and it is actually more complicated than the
equation on the real axis. This difficulty is avoided in the so
called static approximation \cite{zt}, which makes use of the
Euclidean BS amplitude only but brings an additional error
increasing with the momentum transfer.

These problems disappear if one expresses the form factor through the
Minkowski BS amplitude. For the ladder kernel, the BS equation in
Minkowski space was solved in \cite{KW}. For separable
interactions, an approach in Minkowski space was developed and
applied to the nucleon-nucleon system in~\cite{bbmst}. In
\cite{ADT}, the effect of the cross-ladder graphs  in the BS
framework  was estimated with the kernel represented through a
dispersion relation.

Recently, a new general method to find the Minkowski BS amplitude
has been developed \cite{bs1}. This method is valid for any kernel
given by Feynman graphs. In the case of spinless particles, it was
tested for the ladder and cross ladder kernels \cite{bs2}.

Having found the Minkowski BS amplitude, we can calculate EM form
factor without any approximation. This allows us to check the
validity of Wick rotation in the form factor integral, the
accuracy of the static approximation and, in addition, to make a
comparison with light-front dynamics (LFD) calculations. This is
the aim of the present study. A first description of these
results, without any derivation, can be found in \cite{kcm0607}.

This article is organized as follows. In sect. \ref{BSE} we
briefly describe the method \cite{bs1} for solving  the BS
equation in Minkowski space and give new validity tests. In sect.
\ref{minkow} the Minkowski space calculation of form factor is
presented. In sect. \ref{euclid} the analogous Euclidean space
computation is carried out, including its static approximation.
In sect. \ref{LFDapp} the form factor in the LFD framework is
calculated. The comparison of numerical calculations performed in
the different approaches considered in this work is given in sect.
\ref{num}. Finally, some concluding remarks are presented in sect.
\ref{concl}.

\section{The method}\label{BSE}
According to \cite{bs1,bs2}, the Minkowski space BS amplitude is
found in terms of the Nakanishi integral representation
\cite{nakanishi,N_63}:
\begin{eqnarray}\label{bsint}
\Phi_M(k;p)&=&-{i\over \sqrt{4\pi}}\int_{-1}^1dz\int_0^{\infty}d\gamma \nonumber\\
&\times& \frac{g(\gamma,z)}{\left[\gamma+m^2
-\frac{1}{4}M^2-k^2-p\cdot k\; z-i\epsilon\right]^3}.
\end{eqnarray}
The weight function $g(\gamma,z)$ itself is not singular, whereas
the singularities of the BS amplitude are fully reproduced by this
integral. For example, if we set $-{i\over
\sqrt{4\pi}}g(\gamma,z)=1$ and calculate the integral, we find
$$
\Phi_M(k;p)=\frac{i^2}{\left[(\frac{p}{2}+k)^2-m^2+i\epsilon\right]
\left[(\frac{p}{2}-k)^2-m^2+i\epsilon\right]},
$$
{\it i.e.} just the product of two free propagators in (\ref{bs}).
$\Phi_M$ in the form (\ref{bsint}) is substituted into the BS
equation (\ref{bs}) and after some mathematical transformations
\cite{bs1}, one obtains the following integral equation for
$g(\gamma,z)$:
\begin{eqnarray} \label{bsnew}
&&\int_0^{\infty}\frac{g(\gamma',z)d\gamma'}{\Bigl[\gamma'+\gamma
+z^2 m^2+(1-z^2)\kappa^2\Bigr]^2} =
\nonumber\\
&&\int_0^{\infty}d\gamma'\int_{-1}^{1}dz'\;V(\gamma,z;\gamma',z')
g(\gamma',z'),
\end{eqnarray}
where $\kappa^2 = m^2- \frac{1}{4}M^2$ and $V$ is calculated in
terms of kernel $K_M$  \cite{bs1}. A remarkable point  is that
equation  (\ref{bsnew}) is strictly equivalent to the original BS
equation (\ref{bs}) and does not contain any singularity neither
in $g(\gamma,z)$ nor in $V$. Once $g(\gamma,z)$ is found by
solving (\ref{bsnew}), the Minkowski BS amplitude is obtained by
expression (\ref{bsint}).

Equation for the Euclidean BS amplitude is obtained from the
Minkowski one (\ref{bs}) in the center of mass frame $\vec{p}=0$,
by the Wick rotation and substitution $k_0=ik_4$
\begin{eqnarray}
&&\left[\left(k_4^2+k^2+m^2-\frac{M^2}{4}\right)^2+M^2k_4^2\right]
\Phi_E(k_4,\vec{k})= \nonumber \\
&&\int \frac{ dk'_4
d^3k'}{(2\pi)^4}K_E(k_4,\vec{k};k'_4,\vec{k'})\Phi_E(k'_4,\vec{k'}),
\label{bseuc}
\end{eqnarray}
where
\begin{equation}\label{PhiE}
\Phi_E(k_4,\vec{k})=\Phi_M(ik_4,\vec{k};M,\vec{p}=0)
\end{equation}
and
$K_E(k_4,\vec{k};k'_4,\vec{k'})=K_M(ik_4,\vec{k};ik'_4,\vec{k'};M,\vec{p}=0).$
By the method developed in \cite{bs1} we can restore the Euclidean BS
amplitude, {\it i.e.}, find the solution of (\ref{bseuc}), by
setting in (\ref{bsint}) $k_0=ik_4$ and $\vec{p}=0$.

All three equations (\ref{bs}), (\ref{bsnew}) and (\ref{bseuc})
are equivalent to each other in the sense that they give the same
$M^2$. In \cite{bs1} the equivalence of (\ref{bsnew}) to the
initial BS equation (\ref{bs}) has been checked numerically for
the ladder kernel. It was found that the binding energies $B=2m-M$
coincide with high precision.

\begin{figure}[htbp]
\begin{center}
\mbox{\epsfxsize=7.8cm\epsfysize=6.cm\epsffile{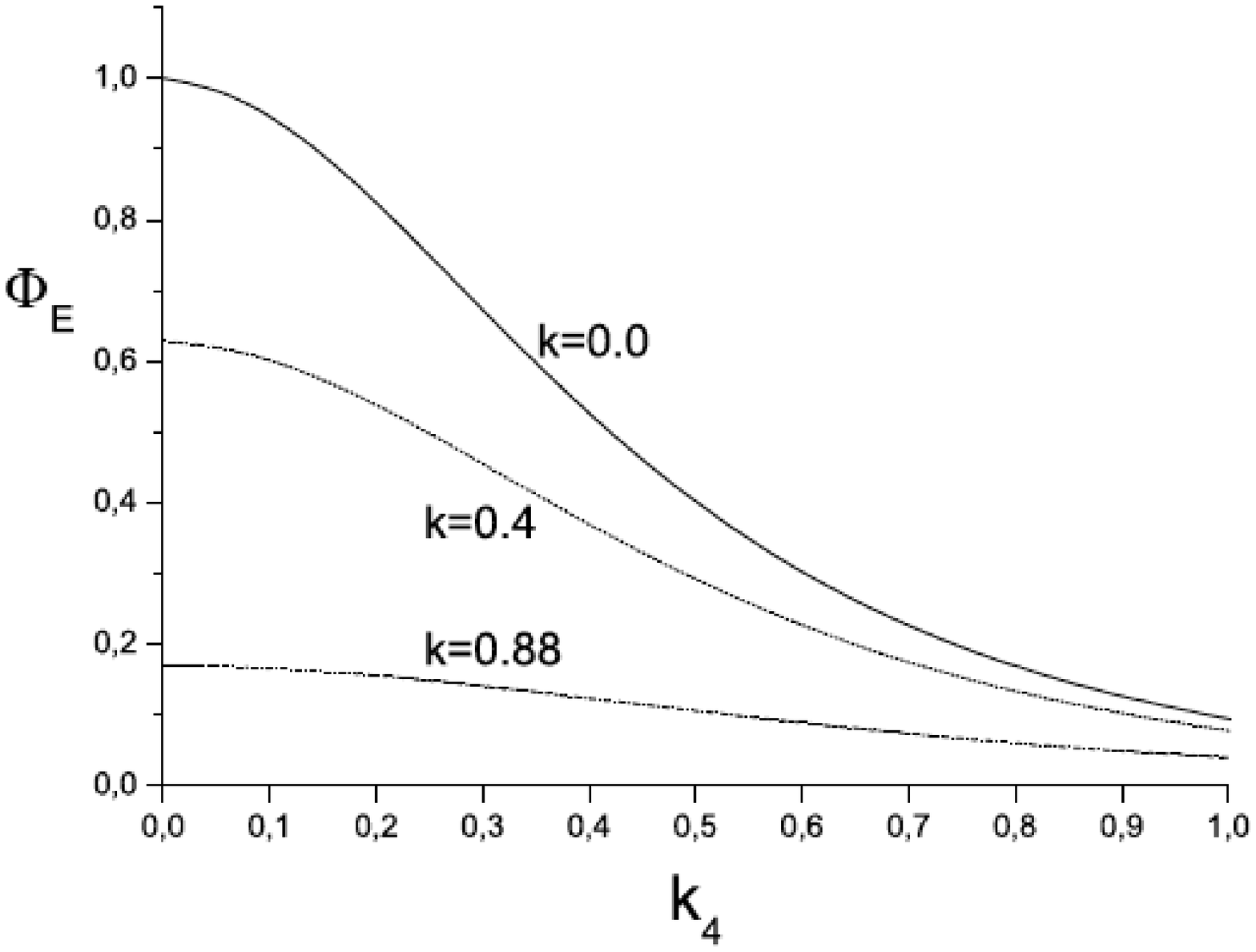}}
\hspace{4mm}
\mbox{\epsfxsize=7.8cm\epsfysize=6.cm\epsffile{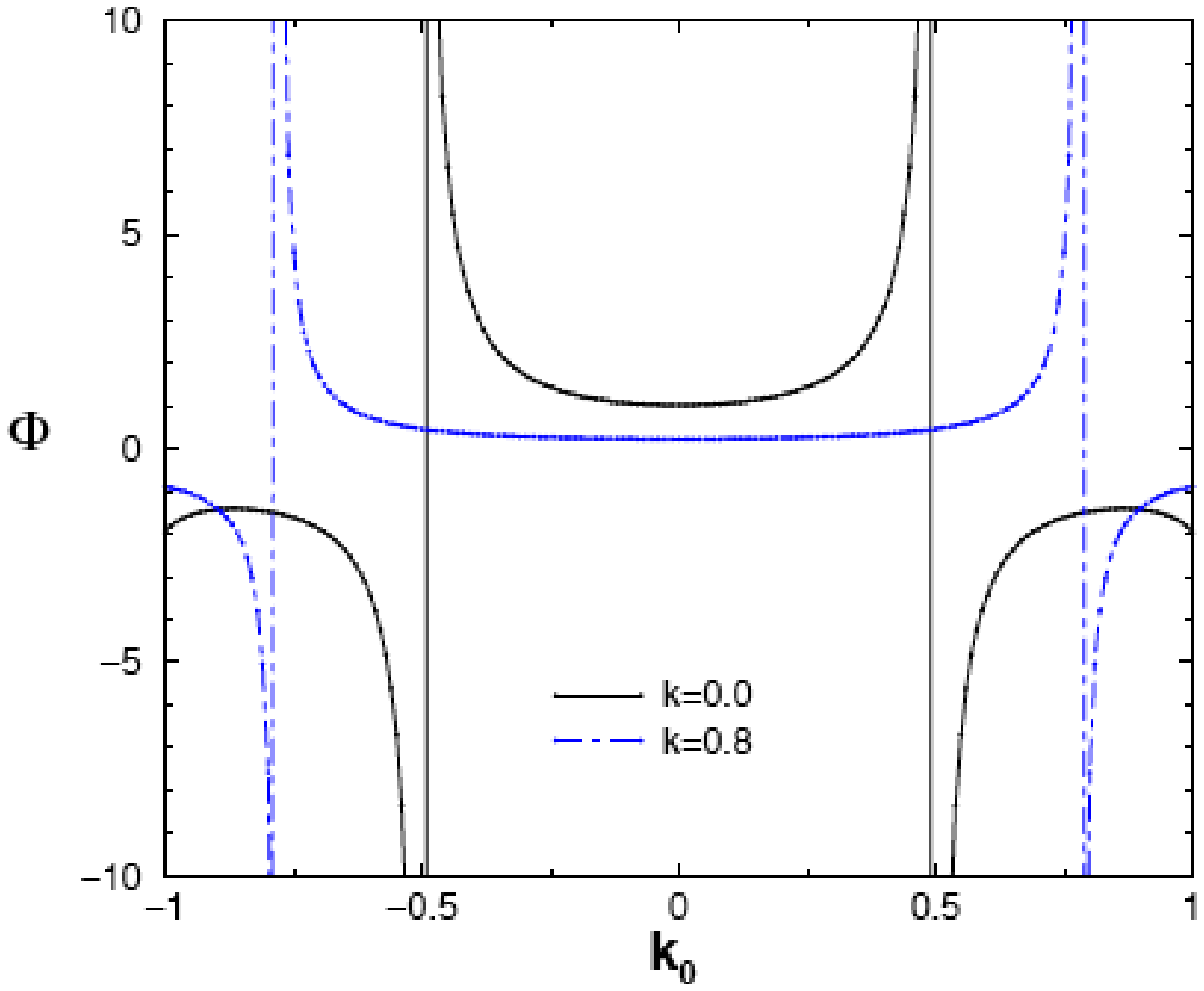}}
\end{center}
\caption{Top: The Euclidean BS amplitude (for ladder kernel)
obtained by (\ref{bsint}) with $k_0=ik_4$, for different values of
$k$. Or (indistinguishable) the one calculated by direct
resolution of BS equation in Euclidean space (\ref{bseuc}). The
exchange mass is $\mu=0.5$.
\newline Bottom: The
corresponding amplitude in Minkowski space, obtained in
\cite{bs1,LC05}. We use the units $m=1$.} \label{fig_psi1}
\end{figure}

For more complete checks, we present here two additional tests. In the
first one, still for the ladder kernel, we compare the Euclidean
BS amplitudes obtained by inserting the solution of (\ref{bsnew})
into (\ref{bsint}), where $k=(ik_4,\vec{k})$, $p=(M,\vec{0})$,
with the one found by directly solving (\ref{bseuc}). These
amplitudes are plotted in fig. \ref{fig_psi1}  (top).  They
coincide at a level better than 0.2\% on the whole range of
$(|\vec{k}|,k_4)$ considered and are indistinguishable from each
other in the graph. On the contrary, the Euclidean BS amplitude
strongly differs from the initial Minkowski one, shown at the
bottom part of fig. \ref{fig_psi1}.  It is worth reminding that so
different amplitudes (compare top and bottom of fig.
\ref{fig_psi1}) correspond to one and the
same mass $M$.

\begin{table*}[htbp][!ht]
\begin{center}
\caption{Coupling constant $\alpha$ for given values of the
binding energy $B$ calculated by the eq. (\ref{bsnew}) and by the
Euclidean BS equation (\ref{bseuc}) for the ladder +cross-ladder
(L+CL) kernel. The exchanged mass: $\mu=0.5$.}
\label{tab1}       
\begin{tabular}{|l|llllll|}
\hline\noalign{\smallskip} $B$ & 0.01 & 0.05 & 0.10 & 0.20 & 0.50
& 1.00
\\
\noalign{\smallskip}\hline\noalign{\smallskip}
 $\alpha$, eq. (\protect{\ref{bsnew}})  &1.206 &1.607 &1.930 &2.416
&3.446 &4.549 \\
$\alpha$, Euclid, eq. (\protect{\ref{bseuc}}) &1.205 &1.608 &1.930
&2.417 &3.448 &4.551 \\
\noalign{\smallskip}\hline
\end{tabular}
\end{center}
\end{table*}

\begin{figure}[h!]
\begin{center}
\begin{minipage}{6.cm}
\mbox{\epsfxsize=6cm\epsffile{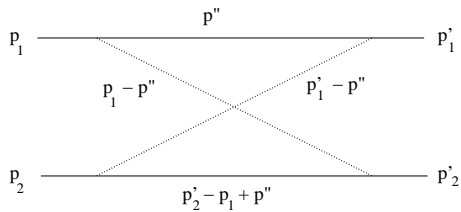}}
\end{minipage}
\end{center}
\caption{Feynman cross ladder graph.\label{CF}}
\end{figure}
The second test incorporates, in addition, the cross ladder kernel
displayed in fig. \ref{CF}. For a massive exchange and a wide
range of binding energies $B$, we carried out precise calculations
(accuracy better than 0.1\%) of the corresponding coupling
constants $\alpha=g^2/(16\pi m^2)$ both by equation (\ref{bsnew})
and by Euclidean equation (\ref{bseuc}) (here $g$ is the coupling
constant in the interaction Hamiltonian). The results are
displayed in the table \ref{tab1} in units of $m$. Their
coincidence demonstrates the validity of both calculations --~in
Minkow\-ski space by the method \cite{bs1} and in Euclidean one.

\begin{figure}[htbp]
\vspace{4mm}
\begin{center}
\mbox{\epsfxsize=7.8cm\epsfysize=6.cm\epsffile{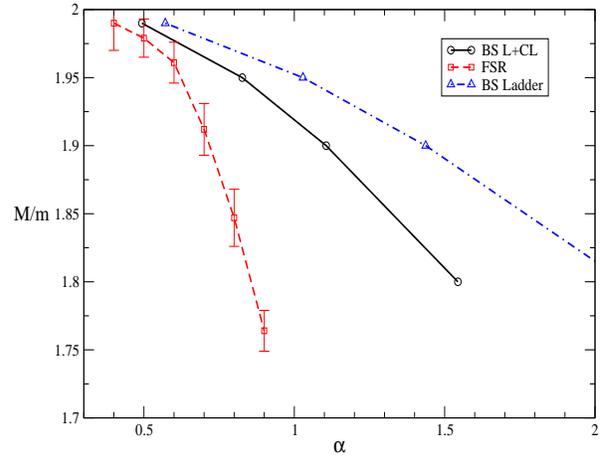}}
\end{center}
\caption{Ground state mass $M$ obtained by the Bethe-Salpeter
ladder and cross-ladder kernel, compared with the
Feynman-Schwin\-ger representation results for an exchanged mass
$\mu=0.15$. } \label{fig_B1}
\end{figure}

The ladder or ladder +cross-ladder kernels are good enough to
check the applicability of the method \cite{bs1}. We would like to
emphasize, however, that both kernels give a rather crude
approximation of the full interaction. Figure \ref{fig_B1} shows
the ground state mass $M$ obtained, for $\mu=0.15$, by the BS
ladder and ladder +cross-ladder kernels together with
Feynman-Schwin\-ger representation results \cite{NT_PRL_96}. The
latter incorporates all the higher order cross box contributions
in the kernel, but not the self energy. Even at low binding
energies, the ladder and cross-ladder results differ by at least
20\%, this difference reaching more than a factor 2 around
$B/m=1$. On the other hand, the results obtained by
Feynman-Schwin\-ger representation departs strongly from the
BS(L+CL) as soon as $B$ is bigger than $0.05\,m$. The cross-ladder
kernel thus gives a non negligible contribution to the total mass
of the system in the right direction, but a larger contribution
remains to be included, due to the higher order terms. Notice
however that the underlying field theory with cubic boson-boson
interaction is unbounded from below \cite{baym}. This instability
(for the interaction $g\phi\chi^2$) appears when infinite number
of the $\chi^2$ loops in the field $\phi$ self energy is included
\cite{Gross_Savkli_Tjon}.

\section{EM form factor via Minkowski BS amplitude}\label{minkow}

\begin{figure}[h!]
\begin{center}
\begin{minipage}{6.cm}
\mbox{\epsfxsize=6cm\epsffile{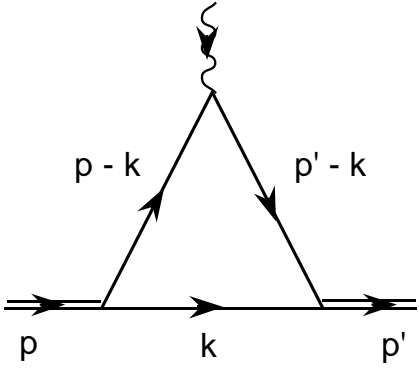}}
\end{minipage}
\end{center}
\caption{Feynman diagram for the EM form factor. \label{triangle}}
\end{figure}
The electromagnetic vertex is shown in
Fig.~\ref{triangle}. We suppose that one of the particles is
charged. By applying the Feynman rules to this graph, we get:
\begin{eqnarray}\label{ffGam}
&&(p+p')^\nu F_M(Q^2)=i\int \frac{d^4k}{(2\pi)^4}\,
\frac{(p+p'-2k)^{\nu}}{(k^2-m^2+i\epsilon)}
\nonumber\\
&&\times\, \frac{\Gamma \left(\frac{1}{2}p -k,p\right)\Gamma
\left(\frac{1}{2}p'-k,p'\right)}{[(p-k)^2-m^2+i\epsilon]
[(p'-k)^2-m^2+i\epsilon]},
\end{eqnarray}
where $\Gamma(k,p)$ is the vertex function, related to the BS
amplitude by:
\begin{equation}\label{Phi}
\Phi_M(k;p)=\frac{\Gamma(k,p)}{\left[(\frac{p}{2}+k)^2-m^2+i\epsilon\right]
\left[(\frac{p}{2}-k)^2-m^2+i\epsilon\right]}.
\end{equation}
Therefore the electromagnetic vertex is expressed in terms of the
BS amplitude by the formula:
\begin{eqnarray}\label{ffbs}
&&(p+p')^{\nu} F_M(Q^2) =i\int \frac{d^4k}{(2\pi)^4}\,
(p+p'-2k)^{\nu}
\nonumber\\
&&\times\;(k^2-m^2)\, \Phi_M \left(\frac{1}{2}p -k;p\right)\Phi_M
\left(\frac{1}{2}p'-k;p'\right).
\end{eqnarray}
We multiply both sides of  (\ref{ffbs}) by $(p+p')_{\mu}$ and
substitute in its r.h.s. the BS amplitude in terms of the integral
(\ref{bsint}). So, the  form factor is given by:
\begin{eqnarray*}
&&(p+p')^2F_M(Q^2)= \int \frac{id^4k}{2(2\pi)^5}\,
 d\gamma dz\,d\gamma'dz'\,g(\gamma,z)g(\gamma',z')
\nonumber\\
&&\times\frac{\left[(p+p')^2-2k\cdot (p+p')\right]\,(m^2-k^2)}
{\left[\gamma+m^2 -\frac{1}{4}M^2-(\frac{1}{2}p-k)^2-p\cdot
(\frac{1}{2}p-k)\; z-i\epsilon\right]^3}
\\
&&\times \frac{1}{ \left[\gamma'+m^2
-\frac{1}{4}M^2-(\frac{1}{2}p'-k)^2-p'\cdot (\frac{1}{2}p'-k)\;
z'-i\epsilon\right]^3} \nonumber
\end{eqnarray*}
To compute this integral, we use the Feynman parametrization:
$$ \frac{1}{a^3b^3}=\int_0^1\frac{30u^2(1-u)^2d u} {\Bigl(a
u+b(1-u)\Bigr)^6}$$
 and then shift the integration variable:
\begin{equation}\label{k1}
k=k_1+\frac{1}{2}(1+z)u\;p+\frac{1}{2}(1+z')(1-u)\;p'.
\end{equation}
Then the integral over $d^4k_1$ has the form
\begin{equation}\label{integ}
\int \frac{\cdots d^4k_1}{(k_1^2-c+i\epsilon)^6},
\end{equation}
where $c$ does not depend on $k_1$. Though the calculation of this
integral by Wick rotation is standard, we explain it here in more
detail, to emphasize the difference with the calculation performed
using  Euclidean BS amplitude, where the Wick rotation cannot be
done (see sect. \ref{euclid} below). The integrand in
(\ref{integ}) does not contain linear terms in $k_{10}$, but only a constant and a quadratic term.
It has four poles at the values $k_{10}= \pm\sqrt{c^2+\vec{k}^2_1}\mp
i\epsilon$. Their positions do not prevent from the
counter-clock-wise rotation of the integration contour. Therefore,
substituting here $k_{10}=ik_4$, we get, for the constant term in $k_{10}$, the following relations:
\begin{eqnarray}\label{ints}
&&\int \frac{d^4k_1}{(k_1^2-c+i\epsilon)^6}= \int
\frac{dk_{10}d^3k_1}{(k^2_{10}-\vec{k}_1^2-c+i\epsilon)^6}
\nonumber\\
&=& \int \frac{idk_{4}d^3k_1}{(k^2_{4}+\vec{k}_1^2+c)^6} =
\int_0^{\infty} \frac{i2\pi^2k^3dk}{(k^2+c)^6}= \frac{2 i\pi^2}{40
c^4}.
\end{eqnarray}
Calculating similarly the integral for the quadratic term
\begin{equation}\label{int2}
\int\frac{k_1^2\;d^4k_1}{(k_1^2-c+i\epsilon)^6}=-\frac{2i\pi^2}{60
c^3},
\end{equation}
we find the following formula, which is exact for a given
$g(\gamma,z)$:
\begin{eqnarray}\label{ffM}
F_M(Q^2)&=&\frac{1}{2^7\pi^3 N_M}\int_0^\infty d\gamma \int_{-1}^1
dz\, g(\gamma,z)
\\
&\times & \int_0^\infty d\gamma' \int_{-1}^1 dz'\,
g(\gamma',z')\int_0^1 du\,u^2(1-u)^2 \frac{f_{num}}{f^4_{den}}
\nonumber
\end{eqnarray}
with
\begin{eqnarray*}
f_{num}&=&(6 \xi-5)m^2 +
           [\gamma' (1 - u) + \gamma u] (3 \xi-2)
\\
&+& 2M^2 \xi(1-\xi) + \frac{1}{4} Q^2 (1 - u) u (1+z) (1+z')
\\
f_{den}&=& m^2 + \gamma' (1 - u) + \gamma u -
            M^2 (1 - \xi) \xi
\\
&+& \frac{1}{4}Q^2 (1 - u) u (1+z) (1+ z'),
\end{eqnarray*}
where $Q^2=-(p-p')^2>0$. To simplify the formula, we use the
notation:
$$
\xi=\frac{1}{2}(1 + z)u + \frac{1}{2}(1+z')(1-u).
$$
We have also introduced in (\ref{ffM}) the normalization factor $N_M$ which
is found from the condition $F_M(0)=1$.

\section{EM form factor via Euclidean BS amplitude}\label{euclid}
Form factor (\ref{ffM}) was calculated using a well justified Wick
rotation in the variable $k_{10}$, defined by (\ref{k1}). As
explained in sect. \ref{BSE}, the Euclidean BS amplitude in the
rest frame $\Phi_E(k_4,\vec{k})$ is obtained from the Minkowski
one (see eq. (\ref{PhiE})) by Wick rotation in the variable $k_0$.
To express the form factor through $\Phi_E(k_4,\vec{k})$, one should
make the Wick rotation, in the variable $k_0$, in integral
(\ref{ffbs}) (for the moment, we ignore the fact that the BS
amplitude in (\ref{ffbs}) is not in the rest frame; we will come
back to this point later). We will show that, in contrast to
integrals (\ref{ints}) and (\ref{int2}), the Wick rotation in
(\ref{ffbs}) cannot be done without crossing
singularities. Therefore the form factor cannot be expressed
through the Euclidean BS amplitude exactly.

It is enough to illustrate this statement in the simplest case, with
$\Gamma(k,p)=1$ and $p'=p=(M,\vec{0})$, {\it i.e.} $Q^2=0$ and
$\nu=0$. Integral (\ref{ffGam}) then turns into:
$$
I=\int \frac{d^4k}{(2\pi)^4}\,
\frac{2i(M-k_0)}{(k^2-m^2+i\epsilon) [(p-k)^2-m^2+i\epsilon]^2}.
$$
The first propagator has poles at
$k_0=\pm\sqrt{m^2+\vec{k}^2}\mp i\epsilon$ and this does not
create any problem, whereas the second factor has poles at:
$$k_0=M\pm\sqrt{m^2+\vec{k}^2}\mp i\epsilon$$
If $\vec{k}^2<M^2-m^2$, both poles are in the r.h.s. half
plane and the pole at $k_0=M-\sqrt{m^2+\vec{k}^2}+i\epsilon$
prevents from the Wick rotation. The exact result for the form factor
should incorporate the residue in this pole and therefore it is
not reduced to the integral obtained from (\ref{ffbs}) by the
naive replacement $k_0=ik_4$. If the residue is omitted (or, in
realistic case, if the contributions of other possible
singularities of $\Gamma(k,p)$ crossed by the rotated contour are
omitted) the result is approximate. In practice, taking into
account the contributions of these unavoidable singularities is
impossible and, hence, the form factor calculated through the
Euclidean BS amplitude is always approximate.

Shifting the variable $k_0$ (for example, $k_0\to
-k_0+\frac{1}{2}p_0$, to transform the argument
$\frac{1}{2}p_0-k_0$ of the BS amplitude in (\ref{ffbs}) into
$k_0$) does not help. The situation remains the same for
non-trivial $\Gamma(k,p)$ and for non-zero $Q^2$.

In addition, there is another reason which does not allow to
express the form factor via Euclidean BS amplitude. The latter is
determined by eq. (\ref{bseuc}) in the rest frame $\vec{p}=0$ and
it is related to the Minkowski one by eq. (\ref{PhiE}). However, the
form factor is expressed through the BS amplitude with non-zero
total momenta $\vec{p}$ and $\vec{p'}$ which due to scattering are
different in initial and final states. Hence, after Wick rotation,
we need to know
\begin{equation}\label{Phibar}
\Phi_{E}^{boost}(k_4,\vec{k};p)=\Phi_M(ik_4,\vec{k};p).
\end{equation}
which differs from the Euclidean BS amplitude
$\Phi_{E}(k_4,\vec{k})$ in eq. (\ref{PhiE}), by non-zero value of
$\vec{p}$. They are identical only at $\vec{p}=0$. The boosted
amplitude $\Phi_{E}^{boost}(k_4,\vec{k};p)$ can be expressed
through $\Phi_{E}(k_4,\vec{k})$, but  only for complex values of its
arguments $k_4,|\vec{k}|$.

Indeed, the Minkowski amplitude $\Phi_M(k_0,\vec{k};p)$ in
r.h.s. of (\ref{Phibar}), for real $k_0$ and for non-zero
$\vec{p}$, can be found from the rest frame amplitude by a boost.
Namely, we can take the BS amplitude
$\Phi_M(k_0,|\vec{k}|;M,\vec{p}=0)$ in the rest frame and
substitute
\begin{eqnarray*}
k_0 & \to & k'_0=\frac{1}{M}(p_0k_0-\vec{p}\cdot\vec{k}),
\\
|\vec{k}| & \to & |\vec{k'}|=\sqrt{{k'}^2_0-k_0^2+\vec{k}^2}.
\end{eqnarray*}
That is:
$$
\Phi_M(k_0,\vec{k};p)=\Phi_M(k'_0,|\vec{k'}|;M,\vec{p}=0).
$$
To get the Euclidean amplitude, we replace here $k_0=ik_4$,
$k'_0=ik'_4$, substitute the result in (\ref{Phibar}) and use the
definition (\ref{PhiE}). Then the relation (\ref{Phibar}) has the
form:
$$
\Phi^{boost}_E(k_4,\vec{k};p)=\Phi_E(k'_4,k')
$$
where $\Phi_E(k'_4,k')$ is the Euclidean BS amplitude in the rest
frame, depending however on the complex arguments:
\begin{equation}\label{kboosted}
k'_4=\frac{1}{M}(p_0k_4+i\vec{p}\cdot\vec{k}),\quad
k'=\sqrt{k_4^2+\vec{k}^2-{k'_4}^2}.
\end{equation}
This requires the knowledge of the Euclidean BS amplitude
$\Phi_E(k'_4,k')$ in the full complex plane. Alternatively, one
can solve the Euclidean BS equation for non-zero $\vec{p}$ (for
real arguments) and obtain $\Phi^{boost}_E(k_4,\vec{k};p)$
directly. These solutions for quark systems were found numerically
in \cite{Maris}. In sect. \ref{euclid1} we will find them, making
the substitution $k_0=ik_4$ in (\ref{bsint}).

In view of these two facts, the EM form factor can be expressed
through the Euclidean BS amplitude only approximately. Below, we
will study the accuracy of the following two approximations.

({\it i}) {\it Naive Euclidean form factor.} In this case, the form factor is found
by the naive substitution $k_0=ik_4$  in the Minkowski expression
(\ref{ffbs}). This corresponds to an approximate Wick rotation which
disregards singularities. However, the BS amplitude in the complex
plane $\Phi_E(k'_4,k')$ can be found exactly, by substituting in eq.
(\ref{bsint}) the complex values (\ref{kboosted}) of boosted
arguments.

({\it ii}) {\it Naive Euclidean form factor in the static
approximation.} In this case, the form factor is still found by
the substitution $k_0=ik_4$ in (\ref{ffbs}). In addition, the
boosted amplitude $\Phi^{boost}_E(k_4,\vec{k};p)=\Phi_E(k'_4,k')$
is approximately replaced by the amplitude at rest
$\Phi^{boost}_E(k_4,\vec{k};M,\vec{p}=0)= \Phi_E(k_4,k)$. Due to
that, the form factor is expressed through the Euclidean BS
amplitude with real arguments.

\subsection{Naive Euclidean form factor}\label{euclid1}

In order to obtain the naive  Euclidean form factor, we start with
the Minkowski space formula (\ref{ffbs}). We use the Breit frame
defined as:
$$
\vec{p'}=-\vec{p},\quad p_0'=p_0=\sqrt{M^2+p^2},\quad Q^2=4p^2.
$$
and shift the integration variable: $ k_0\to -k_0+\frac{1}{2}p_0.$
The spatial
components of eq. (\ref{ffbs}) in the Breit frame are trivially satisfied
(0=0). Taking the time-component, we get:
\begin{eqnarray}\label{ffbsb}
&&2p_0F_M(Q^2)=\int \frac{id^4k}{(2\pi)^4}
\left[\left(\frac{p_0}{2}-k_0\right)^2-\vec{k}^2-m^2\right]
\nonumber\\
&\times&  (p_0+2k_0) \,\Phi_M \left(k_0,\frac{1}{2}\vec{p}
-\vec{k};p\right) \Phi_M \left(k_0,-\frac{1}{2}\vec{p}
-\vec{k};p'\right) \nonumber\\
&&
\end{eqnarray}

We simply replace: $k_0=ik_4$ with real $k_4$, that is,
we neglect the contributions of singularities crossed by the rotated
contour, and we obtain:
\begin{eqnarray}\label{ffbsc}
&&F^{naive}_E(Q^2)=\int \frac{dk_4d^3k}{2p_0(2\pi)^4}\;(p_0+2ik_4)
\\
&&\phantom{F^{naive}_E(Q^2)}
\times\left[m^2+\vec{k}^2-\left(\frac{p_0}{2}-ik_4\right)^2\right]
\nonumber\\
&\times& \Phi^{boost}_E\left(k_4,\frac{1}{2}\vec{p}
-\vec{k};p\right) \Phi^{boost}_E \left(k_4,-\frac{1}{2}\vec{p}
-\vec{k};p'\right) \nonumber
\end{eqnarray}
where $\Phi^{boost}_E(k_4,\vec{k};p)$ is defined in
(\ref{Phibar}). Substituting in r.h.s. of (\ref{Phibar}) the BS
amplitude from eq. (\ref{bsint}), one gets:
\begin{eqnarray}\label{bsint1}
&&\Phi^{boost}_E(k_4,\vec{k},p)=-{i\over \sqrt{4\pi}}\int_{-1}^1dz
\int_0^{\infty}d\gamma \\
&\times& \frac{g(\gamma,z)}{\left[\gamma+m^2
-\frac{1}{4}M^2+k_4^2+\vec{k}^2-(ip_0k_4-\vec{p}\cdot \vec{k})\;
z-i\epsilon\right]^3} \nonumber
\end{eqnarray}
After substituting (\ref{bsint1}) in (\ref{ffbsc}), the form
factor \newline $F^{naive}_E(Q^2)$ is expressed as:
\begin{eqnarray}\label{ffE}
F^{naive}_E(Q^2)&=&\frac{1}{N^{naive}_E}\int_0^\infty d\gamma'
\int_{-1}^1 dz'\; g(\gamma',z') \int_0^\infty d\gamma
\nonumber\\
&\times&\int_{-1}^1 dz\;g(\gamma,z)
\int_0^{\infty}dk_4\int_0^{\infty}d^3k\; f(k_4,\vec{k},\vec{p}),
\nonumber\\
&&
\end{eqnarray}
where
$$
f(k_4,\vec{k},\vec{p})
=-\frac{1}{2^6\pi^5p_0}\mbox{Re}[U(k_4,\vec{k},\vec{p})]
$$

$$
U(k_4,\vec{k},\vec{p})=\frac{\left[k^2+m^2+(k_4+\frac{1}{2}ip_0)^2\right]
(2ik_4+p_0)}{D^3{D'}^3}
$$
with
\begin{eqnarray}\label{A}
D&=&A-\vec{p}\cdot\vec{k}\,(1+z)+\frac{1}{4}(1+2z)\vec{p}^2-i k_4
p_0z-i\epsilon,
\nonumber\\
A&=&\gamma+m^2-\frac{1}{4}M^2+k_4^2+\vec{k}^2
\end{eqnarray}
and $p_0=\sqrt{M^2+\vec{p}^2}$. $D',A'$ are obtained from $D,A$ by
the replacements $\gamma\to\gamma'$, $z\to z'$. The normalization
factor $N^{naive}_E$ is again found from the condition
$F^{naive}_E(0)=1$.

Like Minkowski space form factor (\ref{ffM}), the form factor
(\ref{ffE}) is expressed through the function $g(\gamma,z)$,
satisfying (\ref{bsnew}), but differs from  (\ref{ffM}) by
neglecting singularities crossed when performing the Wick
rotation. Their comparison in sect. \ref{num} will show the error
induced by this approximation.

Note that the denominators $D$ and $D'$, for some momentum
transfer $Q^2$ and the integration variables, may be zero. For
example, for $\gamma=0$, $z=1$, $k_4=0$ we get:
$$
D=m^2-\frac{1}{4}M^2-\frac{1}{16}Q^2+(\vec{p}-\vec{k})^2-i\epsilon
$$
(we used that $Q^2=4p^2$). Since $m^2-\frac{1}{4}M^2$ is positive,
$D$ is always positive too if
\begin{equation}\label{Q2}
Q^2<4(4m^2-M^2).
\end{equation}
If $Q^2>4(4m^2-M^2)$,  $D$ crosses zero for some particular values of
$\vec{p}$ and $\vec{k}$. This singularity is, of
course, integrable (the form factor is always finite), but it is a source of
numerical instability.

\subsection{Naive Euclidean form factor in the static approximation}
As explained in the previous section, the form factor
(\ref{ffbsc}) is expressed
through the BS amplitude $\Phi^{boost}_E(k_4,\vec{k},p)$ which  for $\vec{p}\neq 0$, is in
its turn expressed through the Euclidean
BS amplitude in the complex plane. If we replace the
latter by the Euclidean BS amplitude in the rest frame
$\Phi_E(k_4,\vec{k})= \Phi^{boost}_E(k_4,\vec{k};M,\vec{p}=0)$,
satisfying equation (\ref{bseuc}), we obtain the form factor in the so called static
approximation, which reads:
\begin{eqnarray}\label{ffstat}
F^{stat}_E(Q^2)&=&\frac{1}{N_E}\int \frac{dk_4d^3k}{2(2\pi)^4}
\left(m^2+\vec{k}^2-k_4^2-\frac{1}{4}p_0^2\right)
\nonumber\\
&\times& \Phi_E \left(k_4,\frac{1}{2}\vec{p} -\vec{k}\right)
\Phi_E \left(k_4,-\frac{1}{2}\vec{p} -\vec{k}\right)
\end{eqnarray}
Notice that no approximation in the kinematical factor is done,
though we omit the odd degrees of $k_4$, since after integration
over $dk_4$ they give zero.

As mentioned, $\Phi_E(k_4,\vec{k})$ can be found from equation
(\ref{bseuc}). If $g(\gamma,z)$ is known, $\Phi_E(k_4,\vec{k})$
can be alternatively obtained by equation (\ref{bsint1}) at
$\vec{p}=0$. In this case, the integral  (\ref{bsint1}) is reduced
to:
$$
\Phi_E\left(k_4,\vec{k}\right)=
-i\int_{0}^1dz\int_0^{\infty}d\gamma
\frac{g(\gamma,z)2A(A^2-3k_4^2M^2z^2)}{\sqrt{4\pi}(A^2+k_4^2M^2z^2)^3},
$$
where $A$ is defined in (\ref{A}).

\section{EM form factor via light-front wave  function}\label{LFDapp}
Knowing the Minkowski BS amplitude, we can find the light-front
wave function \cite{cdkm}:
\begin{equation} \label{bs8}
\psi(\vec{k}_{\perp},x) =\frac{(\omega\cdot k_1 )(\omega\cdot k_2
)}{\pi (\omega\cdot p)}\int_{-\infty }^{+\infty }\Phi_M
(k+\beta\omega,p)d\beta.
\end{equation}
Here $\omega$ is a  four-vector with $\omega^2=0$, determining the
orientation of the light-front plane. The
perp-components of vectors, which appear below, are defined
relative to the direction $\vec{\omega}$. Relation (\ref{bs8})
is independent of any model. Substituting (\ref{bsint}) into
(\ref{bs8}), we find the two-body light-front wave function:
\begin{equation}\label{bs10}
\psi(\vec{k}_{\perp},x)
=\frac{1}{\sqrt{4\pi}}\int_0^{\infty}\frac{x(1-x)g(\gamma,1-2x)d\gamma}
{\Bigl(\gamma+\vec{k}_{\perp}^2+m^2-x(1-x)M^2\Bigr)^2}\ .
\end{equation}
The form factor is expressed through this wave function as (see
{\it e.g.} \cite{cdkm}):
\begin{equation}\label{f2b}
F_{LFD}(Q^2)= \frac{1}{(2\pi)^3} \int\psi(\vec{k}_{\perp},x)
\psi(\vec{k}_{\perp}-x\vec{Q}_{\perp},x) \frac{d^2k_{\perp}d
x}{2x(1-x)}, \end{equation} where ${\vec Q}_{\perp}^2=Q^2$.
Substituting in (\ref{f2b}) the wave function
$\psi(\vec{k}_{\perp},x)$ determined by eq.~(\ref{bs10}) and using
the formula
$$ \frac{1}{a^2b^2}=\int_0^1\frac{6u(1-u)d
u}{\Bigl(au+b(1-u)\Bigr)^4},$$ we can easily integrate over
$\vec{k}_{\perp}$ and write the form factor as:
\begin{eqnarray}
&&F_{LFD}(Q^2)=\frac{1}{2^5 \pi^3N_{LFD}}\int_0^{\infty}d\gamma'
\int_0^{\infty}d\gamma \int_0^1 d x\int_0^1 du
\nonumber\\
&&\times \frac{x(1-x)\, u(1-u)\,  g(\gamma,2x-1)g(\gamma',2x-1) }
{\left[u\gamma+(1-u)\gamma'+u(1-u)x^2 Q^2+
m^2-x(1-x)M^2\right]^3}.
\nonumber\\
&& \label{FLFD}
\end{eqnarray}

\section{Numerical results}\label{num}
All the calculations given below have been done with the BS
amplitude found for the ladder+cross ladder kernel. The
constituent mass $m=1$, the exchange mass $\mu=0.5$ and the
coupling constant $\alpha$ has been adjusted to provide the
binding energy $B=1$.
\vspace{4mm}

\begin{figure}[htbp]
\begin{center}
\mbox{\epsfxsize=7.8cm\epsfysize=6.cm
\epsffile{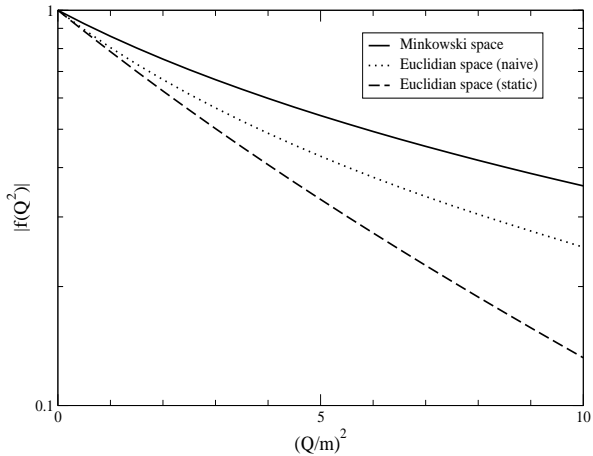}}
\end{center}
\caption{Solid curve: form factor via Minkowski BS amplitude.
Dotted curve: form factor, after Wick rotation, with boosted
Euclidean BS amplitude. Dashed curve: form factor in static
approximation. \label{fig1}}
\end{figure}
For $Q^2 \le 10\, m^2$ ("JLab domain"), the naive Euclidean form
factor and its static approximation are compared with the
Minkowski one in fig. \ref{fig1}. Solid curve is the Minkowski space
calculation, eq. (\ref{ffM}). Dotted curve represents the naive Euclidean form
factor calculated with boosted Euclidean BS amplitude by eq.
(\ref{ffE}). Dashed curve denotes the form factor in the static approximation,
eq. (\ref{ffstat}). The difference between solid and dotted curves
shows that indeed some singularities are missed and, therefore,
the Wick rotation in the variable $k_0$ results in an inaccuracy.
The static approximation (dashed curve) generates an
additional error, increasing with $Q^2$.

Binding energy $B=1$ corresponds to $M=m$. In this case, the
condition (\ref{Q2}) is violated if $Q^2>12\,m^2$. Indeed, our
numerical calculation became unstable if $Q^2$ crosses $\approx
12\,m^2$. That is why the domain of $Q^2$ in fig. \ref{fig1} does
not exceeds $10\,m^2$. This difficulty is absent in the static
approximation.

\begin{figure}[htbp]
\vspace{4mm}

\begin{center}
\mbox{\epsfxsize=7.8cm\epsfysize=6.cm
\epsffile{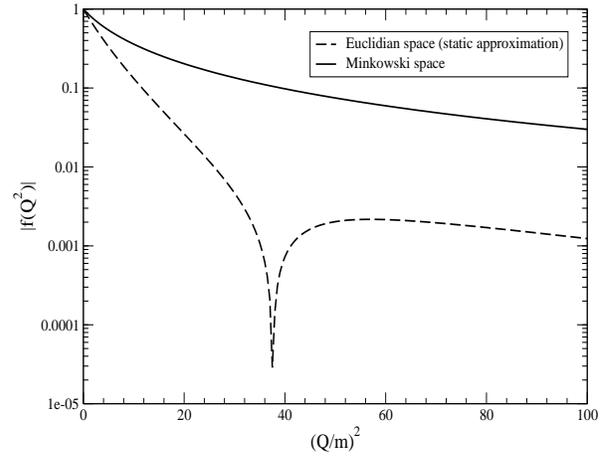}}
\end{center}
\caption{Form factor via Minkowski BS amplitude (solid curve) and
in static approximation (dashed).\label{fig2}}
\end{figure}

Figure \ref{fig2} shows the comparison between form factors
calculated in Minkowski space (solid) and in the static
approximation (dashed) in a wider domain of momentum transfer.
For high $Q^2$, the static form factor is smaller than the
Minkowski one by at least a factor 10. A zero in the static form
factor at $Q^2\approx 38 \, m^2$ is an artefact of the static
approximation, since it is absent in the exact (Minkowski space)
form factor.

\begin{figure}[htbp]
\begin{center}
\mbox{\epsfxsize=7.8cm\epsfysize=6.cm\epsffile{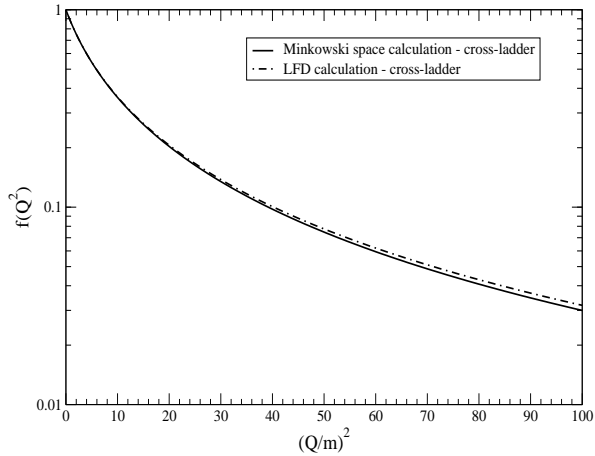}}
\end{center}
\caption{Form factor via Minkowski BS amplitude (solid curve, the
same as in fig. \protect{\ref{fig2}}) and in LFD (dot-dashed
curve). \label{fig3}}
\end{figure}

Figure \ref{fig3} shows the comparison between the form factors
calculated in Minkowski space (solid curve, the same as in fig.
\ref{fig2}) and in LFD, eq. (\ref{FLFD}) (dot-dashed curve). The
LFD form factor is almost indistinguishable from the Minkowski one
in all domain of momentum transfer.

\section{Conclusion}\label{concl}
We have applied the solution of the BS equation in Min\-kow\-ski
space, found by the method developed in \cite{bs1,bs2}, to
calculate the EM form factor and to evaluate the inaccuracy of
different approximations available in the literature. This method gives
the BS amplitudes both in Minkowski and Euclidean spaces as well as
in the full complex plane. We presented two additional validity tests of this
method and demonstrated that it gives the same
Euclidean BS amplitude as the one found by directly solving
equation (\ref{bseuc}). For the ladder and ladder
+cross ladder kernel, it gives the same binding energy.

We calculated the electromagnetic form factor exactly, via Minkowski space BS
amplitude. To express it through the Euclidean solution, one
should carry out the Wick rotation, which, however, requires to
incorporate the  contributions of the singularities, crossed by
the rotating integration contour. In the naive Euclidean form factor,
they are omitted. In addition, after Wick rotation, the Euclidean
BS amplitude in a moving reference frame ({\it i.e.}, boosted BS
amplitude) is expressed through the rest frame one, depending on
 complex arguments. By our method \cite{bs1}, we find the BS
amplitude in complex plane and analyze the error resulting from
naive Wick rotation. The error increases with momentum transfer
and at $Q^2\approx 10 \,m^2$ ("JLab domain") it is about 30\%. In the
static approximation, the error becomes larger, so that at
$Q^2\approx 10 \,m^2$ the Minkowski- and static-approximation form
factors differ by one order of magnitude. The three form factors -- the exact one from
Minkowski BS amplitude, the Euclidean boosted one and in static
approximation -- are found to be close to each other (within a few per cents)
only at relatively small momentum transfer $Q^2\leq m^2$.

The form factors calculated using the Minkowski space BS
amplitude and  the light-front wave function coincide with
each other with very high accuracy. They are almost
indistinguishable.

Note that, in contrast to the Minkowski space BS amplitude, the
light-front wave function is not singular and can be found
directly, from the corresponding 3D equation \cite{cdkm}, without
using any BS formalism and eq. (\ref{bs10}). This advantage,
together with accurate result for the form factor (see figure \ref{fig3}), is one of the
reasons which makes the application of the light-front approach to
the EM form factor rather attractive.

The system of spinless particles considered in this work, provides a simple model giving
a lower limit of different approximations accuracy to the form
factor. One can expect that incorporating spin, these errors would increase.

\section*{Acknowledgement}

One of the authors (V.A.K.) is grateful for the warm hospitality
of the theoretical physics group of the Laboratoire de Physique
Subatomique et Cosmologie, Grenoble, France, where part of the
present work was performed.



\begin{thebibliography}{100}
\bibitem{SB_51} E.E. Salpeter, H.A. Bethe, Phys. Rev. {\bf 84}, 1232 (1951).
\bibitem{nakanishi} N.~Nakanishi, Prog. Theor. Phys. Suppl. {\bf 43}, 1 (1969);
{\bf 95}, 1 (1988).
\bibitem{NT_PRL_96} T.~Nieuwenhuis and J.A.~Tjon, Phys. Rev. Lett.
{\bf 77}, 814 (1996).
\bibitem{W_54} G.C.~Wick, Phys. Rev {\bf 96}, 1124 (1954).
\bibitem{zt}
M.J.~Zuilhof and J.A.~Tjon, Phys. Rev. {\bf 22}, 2369 (1980).

\bibitem{KW} K.~Kusaka, A.G.~Williams, Phys. Rev. D {\bf 51}, 7026 (1995);
K.~Kusaka, K.~Simpson, A.G.~Williams, Phys. Rev. D {\bf 56}, 5071
(1997).

\bibitem{bbmst} S.G.~Bondarenko, V.V.~Burov, A.M.~Molochkov,
G.I.~Smirnov and H.~Toki, Prog. in Part. and Nucl. Phys., {\bf
48}, 449 (2002).

\bibitem{ADT}
A.~Amghar, B.~Desplanques and L.~Theusl, Nucl. Phys. A {\bf 694},
439 (2001).

\bibitem{bs1}
V.A. Karmanov and J. Carbonell,   Eur. Phys. J. A {\bf 27}, 1
(2006) [arXiv:hep-th/0505261].

\bibitem{bs2}
J. Carbonell and V.A. Karmanov,   Eur. Phys. J.  A {\bf 27}, 11
(2006) [arXiv:hep-th/0505262].
\bibitem{kcm0607}
V.A. Karmanov, J. Carbonell, M. Mangin-Brinet, Nucl. Phys. A {\bf
790}, 598c (2007) [arXiv:hep-th/0610158];

V.A. Karmanov, J. Carbonell, M. Mangin-Brinet, {\it Proc. of the
20th Int. Conf. on Few-Body Problems in Physics (FB20), Pisa,
Italy, September 10-14, 2007}, to be published in "Few-Body
Systems" [arXiv:0712.0971].

\bibitem{N_63} N. Nakanishi, Phys. Rev.  {\bf 130}, 1230 (1963);
{\it Graph Theory and Feynman Integrals}, (Gordon and Breach,
New-York, 1971).

\bibitem{LC05}
V.A.~Karmanov and J.~Carbonell,  Nucl. Phys. B (Proc.Suppl.) {\bf
161}, 123 (2006) [arXiv:nucl-th/0510051].

\bibitem{baym} G. Baym, Phys. Rev. {\bf 117}, 886 (1960).

\bibitem{Gross_Savkli_Tjon} F.~Gross, C.~Savkli and J.~Tjon,
Phys. Rev.  D {\bf 64}, 076008 (2001) [arXiv:nucl-th/0102041].

\bibitem{Maris}
   P.~Maris and P.~C.~Tandy,
   Nucl.\ Phys.\ B\ (Proc.\ Suppl.)\  {\bf 161}, 136 (2006)
   [arXiv:nucl-th/0511017];
\\
   M.~S.~Bhagwat and P.~Maris,
   Phys.\ Rev.\  C {\bf 77}, 025203 (2008)
   [arXiv:nucl-th/0612069].


\bibitem{cdkm}
J.~Carbonell, B.~Desplanques, V.A.~Karmanov and\\ J.-F.~Mathiot,
Phys. Rep. {\bf 300}, 215 (1998) [arXiv:nucl-th/9804029].

\end{thebibliography}
\end{document}